\documentclass{emulateapj}
\usepackage{subeqn}
\usepackage{verbatim}

\shorttitle{Phase Coherence of MHD Turbulence}

\shortauthors{BURKHART \& LAZARIAN}
\begin{document}

\title{The Phase Coherence of Interstellar Density Fluctuations}

\author{ Blakesley Burkhart\altaffilmark{1}, A. Lazarian\altaffilmark{2}}
\altaffiltext{1}{Harvard-Smithsonian Center for Astrophysics, 60 Garden St., Cambridge, MA 0213}
\altaffiltext{2}{Astronomy Department, University of Wisconsin, Madison, 475 N. 
Charter St., WI 53711, USA}

\begin{abstract}

Studies of MHD turbulence  often investigate the Fourier power spectrum to provide information on the nature of the turbulence cascade.
However, the Fourier power spectrum only contains the Fourier amplitudes and rejects all information regarding the Fourier phases.
Here we investigate the utility of two statistical diagnostics for recovering information on Fourier  phases in ISM density data: the averaged amplitudes of the bispectrum and the phase coherence index (PCI), a new phase technique for the ISM.  
We create 3D density and 2D column density maps using a set of simulations of  isothermal ideal MHD turbulence with a wide range of sonic and Alfv\'enic Mach numbers.   
We find that the bispectrum averaged along different angles with respect to either the $k_1$ or $k_2$ axis is primarily sensitive to the sonic Mach number while averaging the bispectral amplitudes over different annuli is sensitive to both the sonic and Alfv\'enic Mach numbers.   The PCI of density suggests that the most correlated phases occur in supersonic sub-Alfv\'enic
turbulence and also near the numerical dissipation regime.    This suggests that non-linear interactions with correlated phases are strongest in shock dominated regions, in agreement with findings from the solar wind.
Additionally, our results are particularly encouraging as they suggests the phase information contained in the bispectrum and PCI can be used to find parameters of turbulence in column density maps.

\end{abstract}

\section{Introduction}
\label{intro}
Magnetohydrodynamic (MHD) turbulence is ubiquitous in space plasmas across many orders of magnitude in scale.
This includes the solar wind at astronomical unit scales,  the interstellar medium (ISM) at parsec scales, and intercluster medium (ICM), 
at  megaparsec scales (see for a review  Elmegreen \& Scalo 2004).
In the solar wind, turbulence velocity and magnetic perturbations can be directly measured via in-situ spacecraft
however, in  the ISM and ICM turbulence must be measured indirectly through line of sight tracers such as
column density and spectral lineprofiles (see Lazarian 2009).

The traditional  measure of  turbulence on Earth as well as in astrophysical environments is the spatial Fourier power spectrum.
This is because the turbulence energy transfer
cascade can be studied by examining the Fourier power spectrum and the sources and sinks of energy, 
including the injection and dissipation scales, can be identified. 
For studies of turbulence in the ISM, the power spectrum of density and velocity (and its variants such as
the structure function and delta variance) has been suggested
 by several authors to provide information on the spatial and kinematic scaling of turbulence, 
sonic Mach number and injection/dissipation scales  \citep{Kowal07, Burkhart10, Ossenkopf01, Collins12, Federrath13}.

The power spectrum is defined as:
\begin{equation}
P(\overline{k})=\sum_{\overline{k}=const.}\tilde{F}(\overline{k})\cdot\tilde{F}^{*}(\overline{k})
\end{equation}
where $k$ is the wavenumber and $\tilde{F}(\overline{k})$ is the Fourier transform of the field under study.

Equation 1 demonstrates a critical limitation of the Fourier power spectrum which is
that it contains only the information on amplitudes and disregards all the phase information.
This is problematic for studies of MHD turbulence because interactions among MHD waves
can produce finite correlations of wave phases which are completely missed
by the power spectrum. The coherence or randomness of phases in the MHD turbulence
cascade is of critical importance for particle transport and the understanding of 
wave interaction.  In light of this, several authors have suggested various techniques
which extend beyond the power spectrum to include
information on phases.  

One such technique that preserves the phase information has been applied in the context of the ISM
is the bispectrum. The bispectrum is closely related to the power spectrum.
The Fourier transform of the second-order cumulant,
i.e. the autocorrelation function, is the power 
spectrum while the Fourier transform of the third 
order cumulant is known as the bispectrum. The bispectrum contains information
on both amplitudes and phases and has been applied to both MHD simulations (Burkhart et al. 2009, Cho \& Lazarian 2009)
and observations of neutral hydrogen (Burkhart et al. 2010).   These studies found that the bispectrum
is a sensitive diagnostic for the sonic and Alfv\'enic Mach numbers and could describe  the behavior of 
nonlinear mode correlation across spatial scales.

Another technique, used thus far in the context of the  solar wind community, to investigate  
phase information is the so-called phase coherence index (PCI, see Hada et al. 2003; Koga \& Hada 2003; Koga et al. 2007; Chian et al. 2008; Chian et al. 2010). 
The PCI uses a surrogate data set in which the phase information
in a particular image or signal is randomized and another in which the 
phase information is perfectly correlated.  These surrogate data share the same
power spectrum with each other and the original data  however have different 
phases.  Comparison of the original data and the surrogates gives insight into the 
level of phase coherence or randomness.  PCI has been applied to solar wind
observations and simulations but until now there has not been an
application to MHD turbulence in the context of the ISM or ICM.

In this paper we investigate the application of the bispectrum and PCI
on MHD simulations geared towards ISM observations. 
In principle, the starting point for the definition of both the bispectrum and 
 the phase coherence technique is the Fourier transform, $\tilde{F}$.
We can describe the power by computing $P(\overline{k})$ and the phase
distribution $\phi(\overline{k})=tan^{-1}((Im(\tilde{F})/Re(\tilde{F}))$.

 In particular
we are interested in the physical processes that cause nonlinear
phases to be correlated or uncorrelated, i.e. are phase techniques
sensitive to the amplitude of turbulence fluctuations, the magnetic field 
strength or the sonic Mach number? Most importantly, we seek to 
understand if phase techniques could be useful for observations
of ISM turbulence and focus our analysis on fluctuations in 2D column density maps.
This paper is organized as follows:
in Section 2 we discuss the MHD simulations used for the study of the bispectrum and 
PCI. In Section 3 we discuss our results of a new averaging procedure to
the 2D isocontours of the bispectrum.  In Section 4 we present the first application of the PCI to simulations
of ISM MHD turbulence.  Finally in Section 5 and 6 we discuss our results followed by our conclusions.

\section{Simulations}
\label{sims}

We use the database of 3D numerical simulations
of isothermal compressible (MHD) turbulence with resolution
512$^3$ presented in a number of past works (e.g. Kowal, Lazarian \& Beresnyak 2007; Burkhart et al. 2009;
Burkhart et al. 2013).
We refer to these works for the  details of the numerical set-up and here provide a short overview.

We use the isothermal MHD code detailed in
Cho \& Lazarian (2003) and vary the input values for the
sonic (${M}_s=v/c_s$, where v is the flow velocity and $c_s$ is the sound speed) and Alfv\'enic Mach number ($M_A=v/v_A$, where $v_A$ is the Alfv\'en speed). 
The code is a third-order-accurate ENO scheme which solves the ideal MHD equations in a periodic
box with purely solenoidally driving.  The magnetic field consists
of the uniform background field and a turbulent field, i.e:
B = B$_{ext}$ + b. Initially b = 0.

In total we have 14 simulations at resolution $512^3$.  The simulations have sonic Mach numbers ranging from ${M}_s \approx 0.5-10$.
There are two different magnetic field values used in this investigation: ${M}_A \approx 0.7$ (sub-Alfv\'enic) and ${M}_A \approx 2.0$ (super-Alfv\'enic).

\section{Bispectrum}
\label{bis}

The bispectrum technique characterizes and searches for
nonlinear interactions and departures from Gaussianity, which makes it a useful
technique for studies of MHD turbulence in the context of the ISM and solar wind.
This is because as turbulence eddies evolve they transfer energy from 
large scales to small scales generating a hierarchical turbulence cascade
as  $k_1 + k_2$ interact to form $k_3$.
For incompressible flows, under KolmogorovÕs assumptions,
this can be expressed as $k_1 \approx k_2 = k$ and $k_3 \approx  2k$. 
Nonlinear wave-wave interactions take place more strongly in 
compressible and magnetized flows and in that case
we can have $k_1 \ne k_2$.
The utility of the bispectrum or other three-point statistics is that they can 
characterize nonlinear interactions in both  Fourier amplitude and phase (see Barnett 2002; Masahiro \& Bhuvnesh 2004).

The bispectrum can be defined as:

\begin{equation}
B(|\overline{k_{1}}|,|\overline{k_{2}}|)=\sum_{|\overline{k_{1}}|=const}\sum_{|\overline{k_{2}}|=const}A(|\overline{k_{1}}|)\cdot A(|\overline{k_{2}}|)\cdot {A}^{*}(|\overline{k_{1}}|+|\overline{k_{2}}|)
\label{eq:bispectra}
\end{equation}
\noindent

where $k_{1}$ and $k_{2}$ are 
the wave numbers of two interacting 
waves, and $A(\vec{k})$ is the original
 discrete time series data with finite number
 of elements with $A^{*}(\vec{k})$ representing the 
complex conjugate of $A(\vec{k})$.   We refer to Burkhart et al. (2009) for more details about
the numerical calculation of the bispectrum.  
The final result of our calculation is a 2D isocontour image of bispectral amplitudes as a function
of wave vectors $k_{1}$ and $k_{2}$ (all angular information in the bispectral triangles is averaged out).

 
 The bispectrum  of density and column density was studied in Burkhart et al.  (2009) using 2D isocontour plots.  
 Burkhart et al. (2009)
 found that simulations with  higher sonic Mach number  and an increased magnetic field
 produced more mode correlation across a larger range of scales.
 In particular, Burkhart et al. (2009) found this result applied to both 3D density and observable 2D column density.
 Due to the bispectrum's sensitivity to the sonic Mach number and magnetic field when applied to column density, Burkhart et al. (2010)
 performed a follow-up study on the HI column density map of the SMC in order to constrain the nonlinear interaction of turbulence in atomic
 hydrogen gas. Burkhart et al. (2010) also compared the  bispectrum to other statistical methods for obtaining turbulence parameters. 
 
\begin{figure} 
\begin{center}
\includegraphics[scale=.3]{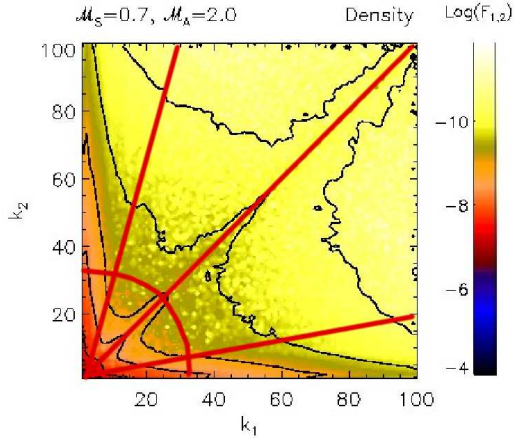}
\caption[ ]{Example of the two  bispectral averaging procedures used in this paper. 
Radial averaging averages all bispectral amplitude values for a given $R^2=k_1^2+k_2^2$ (example of red circular curve that intersects $k_1=k_2=33$).  Angular averaging of the bispectral amplitudes averages all bispectral amplitudes for a given angle  ($\alpha$) as measured from zero (e.g. the three red radial lines shown as an example). }
\label{fig:bic_example} 
\end{center}
\end{figure}

 The issues faced in the above mentioned studies concerning the use of the bispectrum of column density maps for finding
 ${M}_s$ and ${M}_A$ is that the 2D isocontour maps of observations are difficult to compare to simulations.
  In this paper, one of our aims to condense the information provided by the bispectral amplitudes
 into a more readily understandable 1D form. In order to achieve this, we focus on two different averaging procedures
 in order  to distill the information in the 2D isocontour images of the bispectral amplitudes into a 1D plot.
 We use both 
 averaging along different annuli for a given  $R^2=k_1^2+k_2^2$,  and angular averaging of all values along a line with  given angle $\alpha$, as measured from
 the $k_1$ axis\footnote{Note that it doesn't matter if we average with respect to the  $k_1$ or $k_2$ axis since the bispectral amplitudes are symmetric about $k_1$=$k_2$.}. Figure \ref{fig:bic_example} shows an example of both averaging procedures on a 2D isocontour map of the bispectrum amplitudes.
 
\begin{figure*} 
  \begin{center}
\includegraphics[scale=.7]{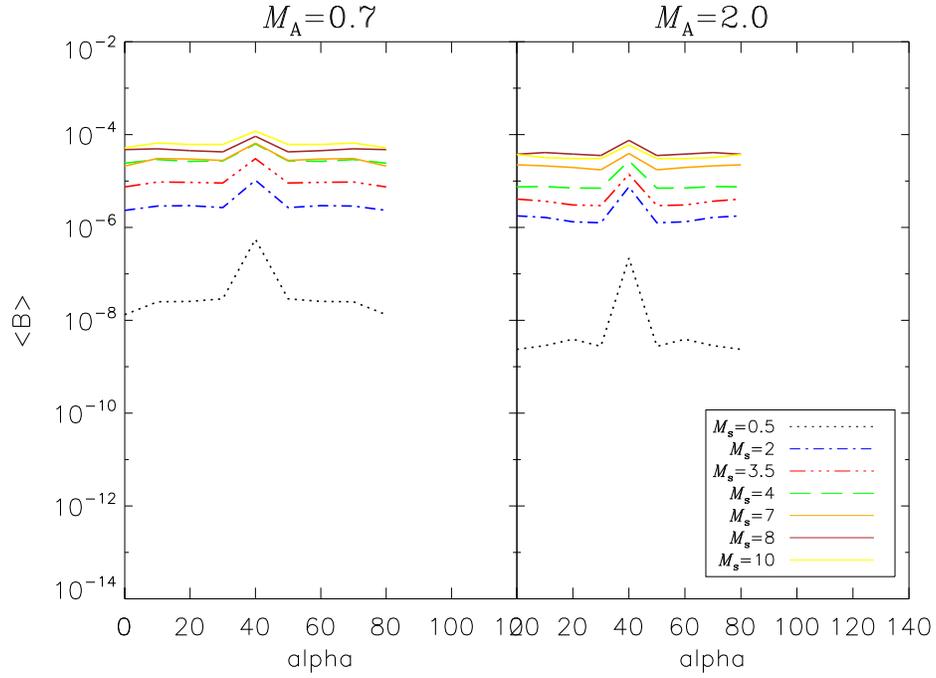}
\caption[ ]{Bispectral amplitudes  of column density along the magnetic field (X-direction) which are radial averaged along different angles ($\alpha$).  }
\label{fig:biccdx_alpha} \end{center} \end{figure*}

 Figure \ref{fig:biccdx_alpha} 
 shows the averaged bispectral amplitudes of column density with LOS in the X direction with an angular averaging of all values along a line with  given angle $\alpha$. We do not find the averaging procedure is sensitive to the LOS direction and show only the direction parallel to the mean magnetic field.  
As expected, past $\alpha=45$ degrees the averaged values of the isocontours are the same as the bispectrum is symmetric about the $k_1$=$k_2$ axis.  Furthermore also as expected the highest amplitude occurs at the averaging along  $k_1$=$k_2$ line, which is equivalent to setting  $\alpha=45$ degrees.

Burkhart et al. (2009) found increased bispectral amplitudes for simulations with higher sonic Mach number, since these simulations depart strongly from Gaussian distributions.  Figure  \ref{fig:biccdx_alpha} reflects this finding in a more compact way.  The higher the sonic Mach number, the larger the bispectral amplitudes are and therefore the higher the average is.   This is true regardless if the simulations are super-Alfv\'enic or sub-Alfv\'enic and we do not see a strong difference between the left and right panels of Figure  \ref{fig:biccdx_alpha} which show different magnetic field strengths.     Thus we can conclude that angular averaging of bispectral isocontour values along a given $\alpha$ is sensitive to the sonic Mach number regardless of the Magnetic field strength.

\begin{figure*} 
  \begin{center}
\includegraphics[scale=.7]{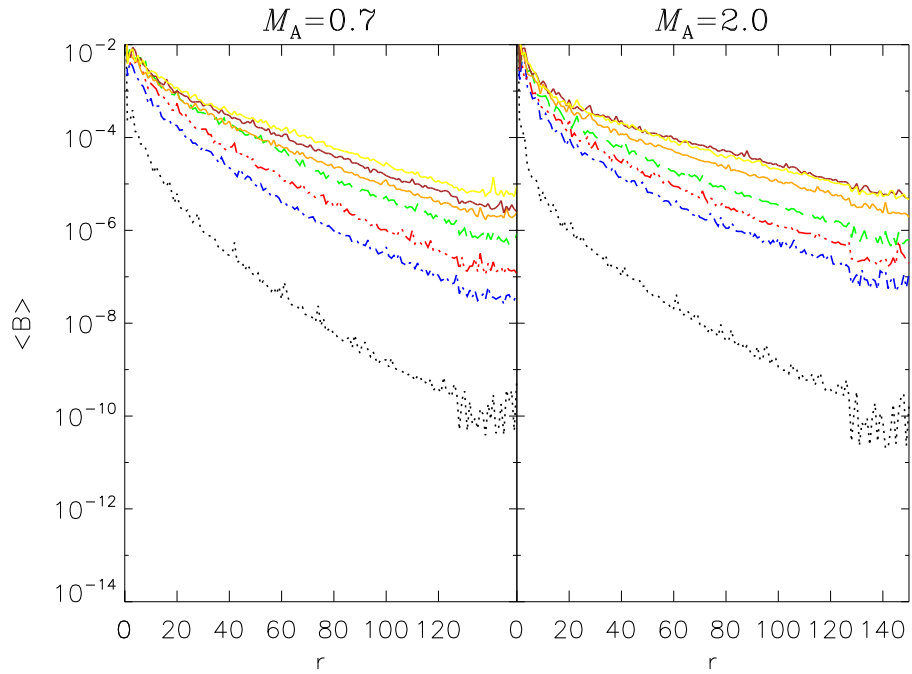}
\caption[ ]{Bispectral amplitudes of column density along the magnetic field (X-direction) averaged along different annuli, i.e.  for a given $r^2=k_1^2+k_2^2$.  The left panel shows sub-Alfv\'enic simulations and the right panel shows super-Alfv\'enic simulations. The color
scheme is the same as in  Figure \ref{fig:biccdx}.}
\label{fig:biccdx} \end{center} \end{figure*}

 Figure \ref{fig:biccdx}  shows the averaged bispectral amplitudes of column density with LOS in the X direction, respectively, with averaging along different annuli for a given  $R^2=k_1^2+k_2^2$.   Similar to the averaging over different $\alpha$, averaging the bispectral amplitudes over different annuli shows a strong sensitivity to the sonic Mach number.  Column density maps produced from simulations with larger sonic Mach number show higher values of averaged bispectral amplitudes.  However, comparing the left and right panels in Figure \ref{fig:biccdx},  it is clear that the averaging along different annuli also shows a sensitivity to the magnetic field strength.  Simulations with ${M}_A=2.0$ 
show slightly larger averaged amplitudes as compared with simulations with ${M}_A=0.7$ for a large range of different sonic Mach numbers.  These findings persist regardless of the LOS direction and we do not find that this diagnostic is very sensitive to the LOS with respect to the mean magnetic field orientation, hence we show only the LOS  parallel to the mean field direction.  
Thus we can conclude that averaging the bispectral amplitudes over different annuli is sensitive to both the sonic and Alfv\'enic Mach numbers.
 
It is particularly encouraging that both averaging procedures have slightly different sensitivities to the parameters of the turbulence.  The annuli averaging is sensitivity to both sonic and Alfv\'enic Mach numbers while the angular averaging is sensitive only to sonic Mach number.
In both cases, the LOS does not seem to play a significant role in the overall amplitudes.  These differences will allow researchers to break the degeneracy in the bispectrum's sensitivity to multiple turbulence parameters.

\section{The Phase Coherence}
\label{phase}

\subsection{Phase Coherence Technique}

\begin{figure*} 
\begin{center}
\includegraphics[scale=.8]{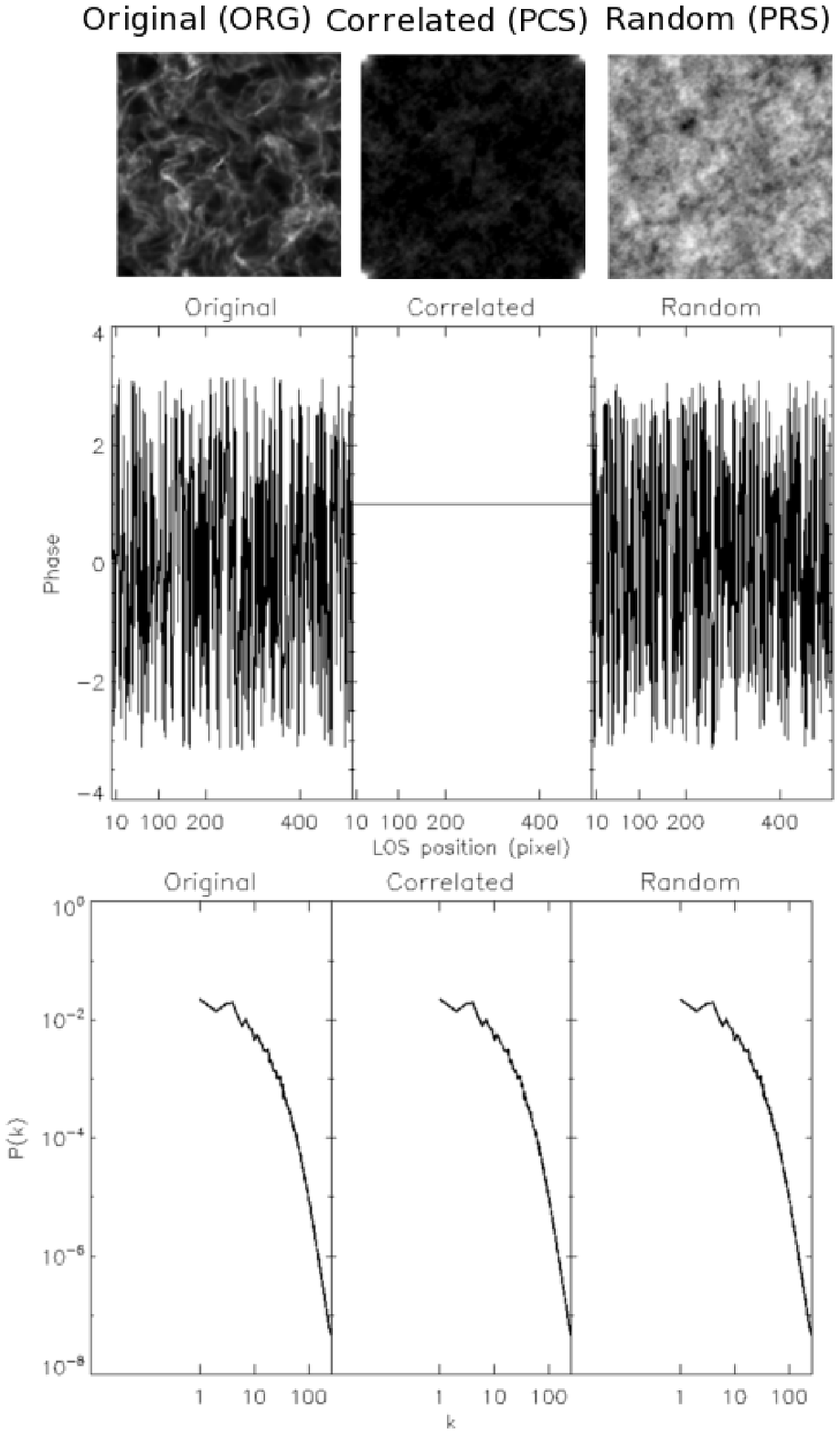}
\caption[ ]{The three data sets required for the application of the phase coherence technique.  The original data image (ORG)  from a supersonic sub-Alfven\'ic is shown in the top left.  From the original data set we construct two surrogate data sets with correlated phase (PCS data, center top image) and with random phases (PRS, top right image).  The same positional line of sight cut through the phase is shown for each data set in the central panel.  The corresponding power spectrum (shown in the bottom panels) is unchanged and the same for all data sets.}
\label{fig:phase_ex} 
\end{center}
\end{figure*}

The PCI was first introduced in Hada et al. (2003) and Koga \& Hada (2003) in order to  evaluate the degree of phase coherence among Fourier modes.  In essence the technique involves the construction of two surrogate data sets from an original N-dimensional data set.
In particular, given an original sequence of data, henceforth denoted as ORG (e.g. time series data or fluctuations of density along a position axis) we can construct two surrogate data sets with the same power spectrum but randomly shuffle the phases for one  data set and, for the other surrogate data set,  perfectly correlate the phases by setting them equal.   We denote the surrogate data with Gaussian random phases as the phase-randomized surrogate (PRS) and the data with correlated phases as the phase-correlated surrogate (PCS).  The three data sets, ORG, PRS and PCS, share exactly the same power spectrum, while their phase distributions are all different.  Figure \ref{fig:phase_ex} shows an example of three different data sets, original plus two surrogate data sets, their identical power spectrum and different phase distributions.

We define the path length similar to that of Koga et al. 2008

\begin{equation}
L(\tau)=\sum_t |x(t+\tau)-x(t)|
\label{eq:path}
\end{equation}

Where $\tau$ is the lag value. In our application of L($\tau$), we use spatial increments whereas Koga et al. (2008) used time series data. When the phases of a given data set are 
correlated the path length will be smaller than in situations where the phases
are randomized.  Because of this 
the phase coherence index can be defined  as:
\begin{equation}
C_\phi(\tau)=\frac{L_{PRS}(\tau)-L_{ORG}(\tau)}{L_{PRS}(\tau)-L_{PCS}(\tau)}
\label{eq:co}
\end{equation}
and gives an evaluation on the degree of coherence in the phases.
If the original data have randomized phases than
C$_\phi$  should be roughly zero.  If C$_\phi$  is close to unity then this indicates that the phases
are nearly completely correlated.

In practice we averaged together five random Gaussian realizations to construct the phases of the PRS data set.  Along with the original data 
(ORG), we use one realization
of the PCS (all phases are set to unity).  We calculate the first order structure function of the ORG, PRS and PCS data sets
(i.e. we apply Equation \ref{eq:path}) and then compute the phase coherence given in Equation \ref{eq:co}.
A schematic of the ORG, PCS and PRS data sets is shown in Figure \ref{fig:phase_ex}.

\subsection{The Phase Coherence of Density}

\begin{figure*} \begin{center}
\includegraphics[scale=.8]{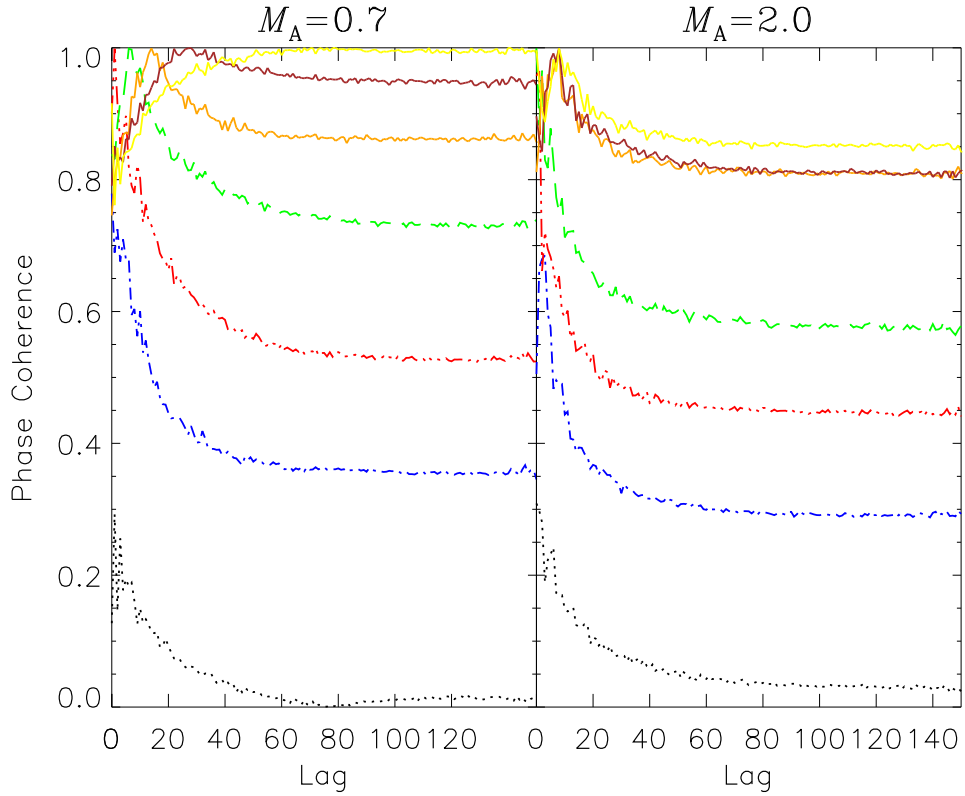}
\caption[ ]{ The phase coherence $C_\phi(\tau)$ vs. $\tau$ of density.  The left panel shows simulations
with sub-Alfv\'enic turbulence while the right panel shows simulations with super-Alfv\'enic turbulence.  Different
sonic Mach number runs are denoted with different colors and linestyles. The color scheme is the same as in Figure \ref{fig:biccdx_alpha}.   }
\label{fig:dncc} \end{center} \end{figure*}

We plot the phase coherence, $C_\phi(\tau)$ vs. $\tau$, of density
in Figure \ref{fig:dncc}.  The left panel shows simulations
with sub-Alfv\'enic turbulence while the right panel shows simulations with super-Alfv\'enic turbulence.  Different
sonic Mach number runs are denoted with different colors and linestyles. 

The phase coherence is peaked at unity for simulations with highly correlated phases and approaches zero when the phases
are random.  The largest values of the phase coherence occur at the smallest lag values and decrease and eventually saturate at higher lag
values.  The saturation occurs roughly at lag=40 for all simulated boxes with smaller values than this being well within the dissipation range of our simulations.  This indicates that the dissipation region of the simulations show the greatest enhancement of phase coherence. 
In the inertial range of the simulations the phase coherence is roughly constant across different lag values.  This is in contrast to the power spectrum (or the analogous structure function), which decrease as a powerlaw in the inertial range but lacks any phase information.

Like other phase analysis techniques such as the bispectrum, $C_\phi(\tau)$ has a strong dependence
on the sonic Mach number of the simulation.  Subsonic simulations (shown with black dotted lines) have phase distributions
that are closer to random ($C_\phi(\tau) \le 0.35$).  This is not surprising since subsonic turbulence has many statistical features in common with a Gaussian distribution, i.e. low bispectral amplitudes (see Burkhart et al. 2009), similar topological features (see Chepurnov et al. 2008), and a lognormal PDF (see Kowal, Lazarian \& Beresnyak 2007). As the sonic Mach number increase the phase coherence also increases.  In the inertial range there is nearly no overlap between different MHD runs for different sonic Mach number (for the same ${M}_A$), making the phase coherence a sensitive diagnostic of the sonic Mach number.  

Comparison between the left and right panel of Figure \ref{fig:dncc} also shows that, like the bispectrum, the phase coherence is also sensitive to the magnetic field information present in the data.  Simulations with a higher mean magnetic field (i.e. sub-Alfv\'enic simulations) show enhanced values of $C_\phi$.  This suggests that the phase coherence might also be used to assess the Alfv\'enic nature of the gas when the sonic Mach number is known.

 \begin{figure*} \begin{center}
\includegraphics[scale=.8]{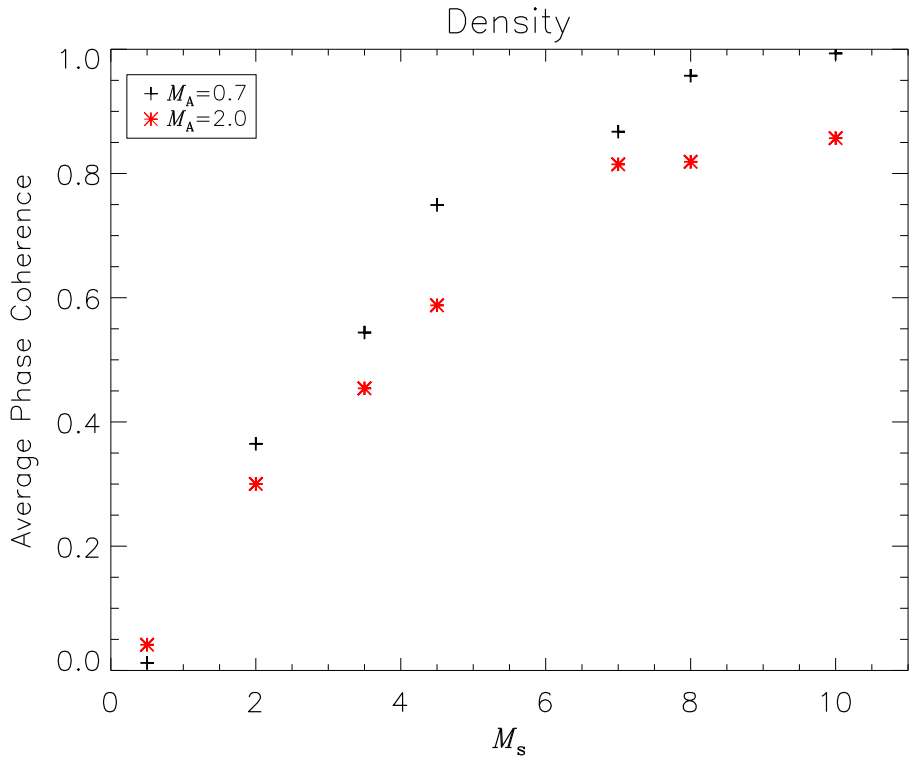}
\caption[ ]{Averaged values of $C_\phi$  vs. the sonic Mach number for 3D density fields. We average values of $C_\phi$ over lag values from 40 to 100. Super-Alfv\'enic simulations are denoted with red stars while sub-Alfv\'enic simulations are denoted with black plus symbols.}
\label{fig:dn_av} \end{center} \end{figure*}

To demonstrate the above point we average the values of $C_\phi$ over lag values from 40 to 100 and plot the average $C_\phi$ vs. sonic Mach number of the simulation in Figure \ref{fig:dn_av}.
The average phase coherence increases with increasing sonic Mach number and levels off near  $C_\phi(\tau)=1$ for Mach numbers greater than 8.  This suggests that the phase coherence will be a sensitive diagnostic of sonic Mach number out to ${M}_s \approx 8$ however for larger sonic Mach numbers the phase coherence approaches unity.   This is similar to the findings of Koga et al. (2008), who found higher phase coherence values at the Earth's bow shock.
Additionally, there is a slight degeneracy with the strength of the magnetic field, which becomes even more apparent when averaging the phase coherence.  Stronger magnetic field produces density fluctuations which have  more correlated phases.



 \begin{figure*} \begin{center}
\includegraphics[scale=.8]{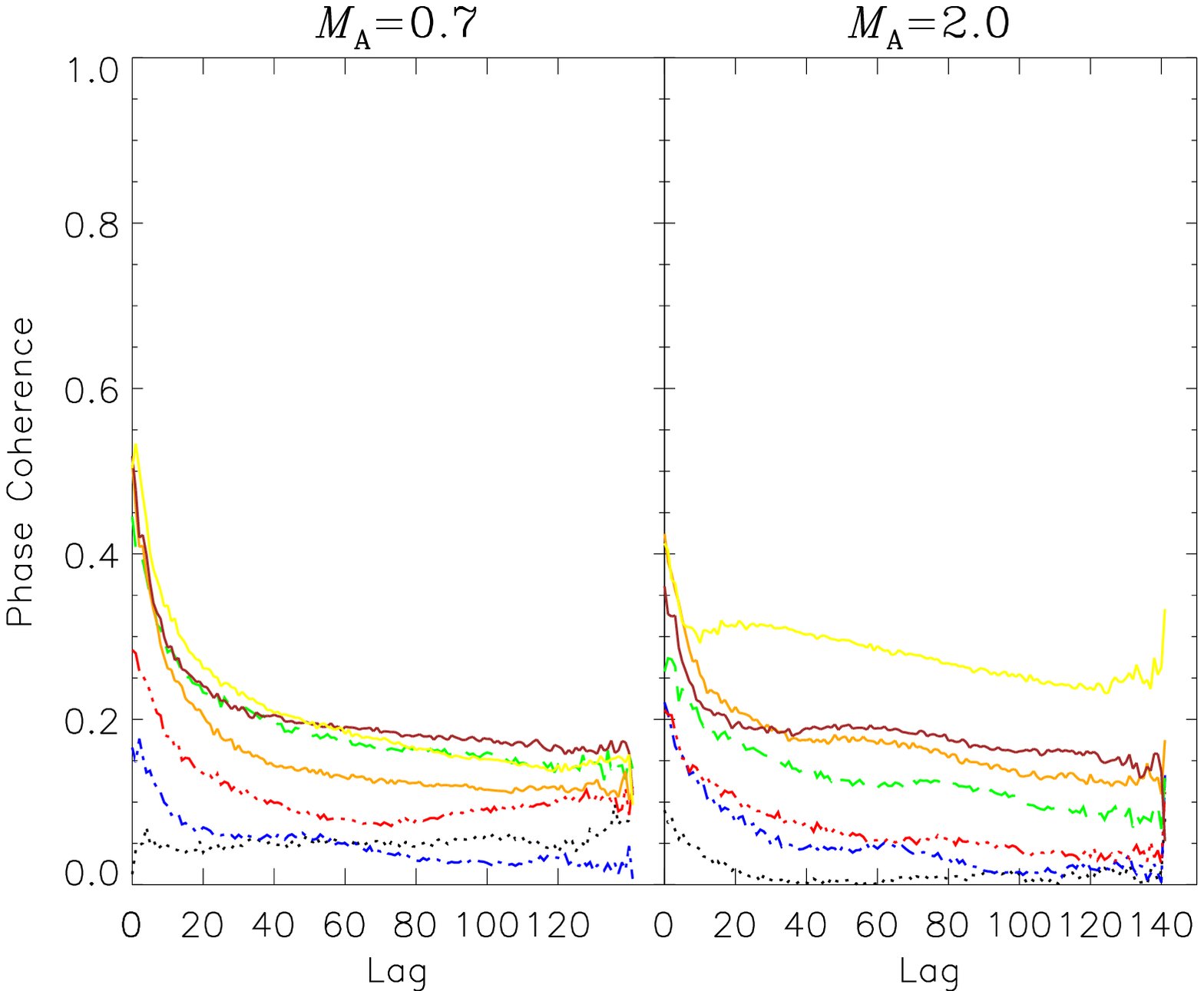}
\caption[ ]{Same as Figure \ref{fig:dncc} but for column density integrated along the Z axis.
The left panel shows simulations with sub-Alfv\'enic turbulence while the right panel shows simulations with super-Alfv\'enic turbulence.  Different sonic Mach number runs are denoted with different colors and linestyles. The color scheme is the same as in Figure \ref{fig:biccdx_alpha}. }
\label{fig:cdz_cc} \end{center} \end{figure*}

 \begin{figure*} \begin{center}
\includegraphics[scale=.8]{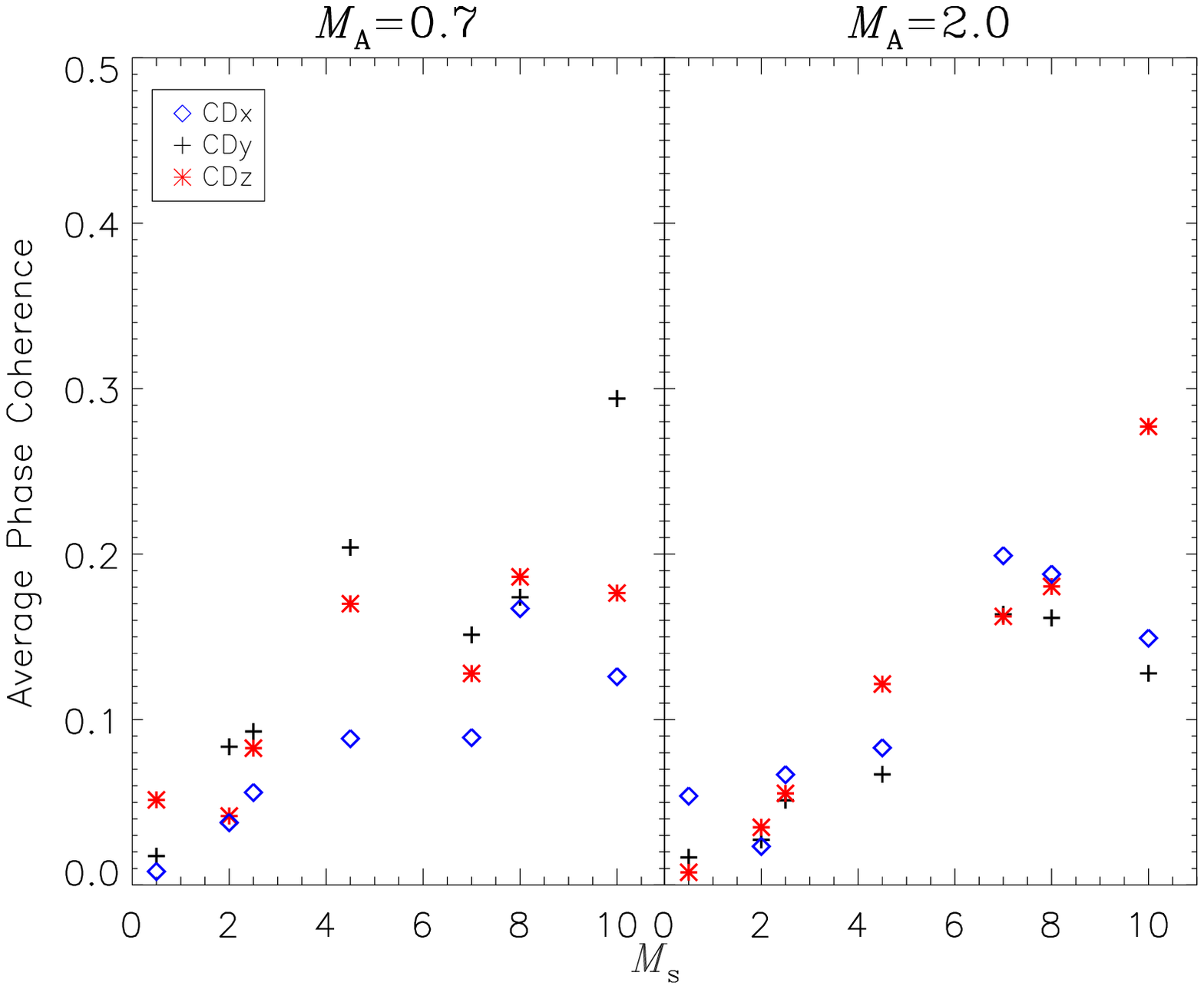}
\caption[ ]{Averaged values of $C_\phi$  vs. the sonic Mach number for 2D column density fields. We average values of $C_\phi$ over lag values from 40 to 100. }
\label{fig:cd_av} \end{center} \end{figure*}

\subsection{The Phase Coherence of Column Density}

We apply the phase coherence to 2D images of column density maps created from the 3D density distributions in order to test the applicability of the PCI for observations of the ISM.  We plot the PCI vs. lag for column density in Figure \ref{fig:cdz_cc}.
Similar to our analysis of the bispectrum we test three different LOS orientations but find that the PCI is not particularly sensitive to the LOS orientation and opt to show the figure only for the perpendicular direction along the Z-axis.

The PCI of the column density shows similar overall trends as that of the 3D density.  In particular, the PCI is very sensitive to the sonic Mach number of the gas, with more supersonic runs having a higher value of the PCI, suggesting that the phases are more correlated overall.   Similarly the dissipation range of the simulation is clearly visible in the overall behavior of the PCI of column density.  This suggests that the PCI might be used to determine the dissipation range in observations that resolve it.
However there are a number of notable differences of the PCI of column density as compared with 3D density.  For example the overall values of the PCI are much lower for column density than they are for density. This is perhaps due to averaging multiple random cells along the LOS to produce the column density maps.
Another major difference is that the PCI's sensitivity to the magnetic field intensity is not as apparent in the column density as it is in density maps.

When we plot the average the values of $C_\phi$ over lag values from 40 to 100 (shown in Figure \ref{fig:cd_av}) for column density using all three LOS orientations the lack of sensitivity of the column density PCI to magnetic field and LOS become even more apparent.  Nevertheless, the PCI of column density is still sensitive to the overall sonic Mach number of the gas and the dissipation scale , making this technique useful for determining these parameters of ISM turbulence from column density maps.

\section{Discussion}
\label{disc}

Phase information is often ignored in studies of turbulence, which generally focus on the Fourier power spectrum
i.e., the Fourier amplitudes.  The studies of power spectrum resulted in substantial progress of our understanding of
interstellar turbulence. For instance, the Big Power Law in the Sky (Armstrong et al. 1995, Chepurnov \& Lazarian 2010),
as well as studies of random densities (see Elmegreen \& Scalo 2004, for a review) and velocities (see Chepurnov et al. 2015, Lazarian 2009 for a review) provided convincing evidence for the existence and importance of turbulence in the interstellar media. 

Nevertheless, power spectra as well as its real space counterparts, namely, structure functions and correlation functions
(see Monin \& Yaglom 1972) cannot provide the full description of the turbulent field. This induced extensive studies
of alternative measures of turbulence. Probability distribution functions of column densities
including different measures obtained with them, e.g. skewness and kurtosis (Burkhart et al. 2009, 2010), Tsalis statistics (Esquivel \& Lazarian 2010, Toffelmire et al. 2011), dispersions (Burkhart \& Lazarian 2012), were considered as tools in order to obtain the properties of turbulence from observations. Together with the techniques for anisotropy of turbulence studies (Lazarian et al. 2002, Esquivel \& Lazarian 2005, 2011, Heyer et al. 2008, Burkhart et. al. 2014) and intermittency studies (Padoan et al. 2004, Kowal et al. 2007) those present an impressive toolbox for quantitative studies of interstellar turbulence. 

At the same time, it is known that  nonlinear interactions among MHD waves are likely to produce finite correlation of the wave phases and therefore the phases should be studied. 
Recently several studies have promoted the study of phase information for ISM MHD turbulence.
These include three-point statistics, such as the bispectrum studied here as well as in Burkhart et al. (2009, 2010).
Furthermore, it was recently shown in Correia et al. (2015) that the principle component analysis (PCA) (see Heyer \&  Schloerb 1997, Heyer et al. 2008).
has sensitivity to the phase information. Intermittency studies also utilize the information of phases.

In spite of the aforementioned studies, this is the first paper, as far as we are aware of, which is entirely focused on 
making use of the information about phases to provide simple measures that can be used to study interstellar turbulence. 
For instance, the bispectrum is a rich measure which presents a two dimensional distribution. In this paper we condensed
the information provided by the bispectrum by presenting two simple functions, which describe the radial and azimuthally averaged
measures of bispectrum.  In addition, we presented the measures of the phase coherence index, which can provide another
way to characterize the phase information related to turbulence. 
We find that in shock dominated turbulence the PCI approaches unity, in agreement with solar wind studies such 
as those of Koga et al. (2008).   This is equally true for the bispectrum, where correlation of wave modes
increases for supersonic highly magnetized turbulence, as first discussed in Burkhart et al. (2009) for the ISM.

Our present paper continues the trend of bringing the statistical techniques that were developed in other areas into studies
of interstellar turbulence. For instance, PCI, similar to Tsallis statistics was first used for the solar wind studies (Burlaga et al
2006, 2007, 2009).
With the advent of new telescopes and precision measurements we expect many of the interstellar turbulence techniques to
be used for studying turbulence in galaxy clusters. In fact, we believe in cross pollination of different branches of turbulence research. For instance, we think that our suggestions of bispectrum averaging may be useful for studying turbulence beyond
the interstellar turbulence domain and could also  be used for cosmological studies, which already employee the bispectrum.


\section{Conclusions}
\label{con}

We investigated the utility of two statistical diagnostics for recovering information on Fourier  phases  in the ISM using a set of simulations of MHD turbulence with a larger range of sonic and Alfv\'enic Mach numbers.   In particular, we focused our study on a new averaging procedure for the bispectrum isocontour amplitudes in order to distill the information in the isocontours into a 1D form. We also introduce the phase coherence index, a new technique for studies of density fluctuations in the ISM.

We  find that:

\begin{itemize}
\item The bispectrum averaged along different angles with respect to either the $k_1$ or $k_2$ axis is primarily sensitive to the sonic Mach number.
\item Averaging the bispectral amplitudes over different annuli is sensitive to both the sonic and Alfv\'enic Mach numbers.  
\item Higher sonic Mach number and larger magnetic field produce density structures which have more correlated phases behavior compared to a random Gaussian distribution of phases.
\item We find that in shock dominated turbulence the PCI approaches unity, in agreement with solar wind studies.
\item The PCI of density  is sensitive to both the sonic and Alfv\'enic Mach numbers. However when applied to column density maps the PCI is sensitive only to the sonic Mach number.
\end{itemize}

\acknowledgments
The research of B.B. is supported by the NASA Einstein Postdoctoral Fellowship.

\end{document}